\documentclass[conference]{IEEEtran}
\IEEEoverridecommandlockouts
\usepackage{cite}
\usepackage{amsmath,amssymb,amsfonts}
\usepackage{amsthm}
\usepackage{booktabs}
\usepackage{array}
\usepackage{listings}
\usepackage[ruled,vlined,linesnumbered]{algorithm2e}
\usepackage{tikz}
\usetikzlibrary{arrows.meta,positioning,fit,backgrounds}
\usepackage{scratch3}
\usepackage{xcolor}
\usepackage{url}
\usepackage[hidelinks]{hyperref}
\usepackage{balance}

\newcommand{\lplayer}{$L_{\mathrm{player}}$}
\newcommand{\lfinal}{$L_{\mathrm{final}}$}
\newcommand{\ldebug}{$L_{\mathrm{debug}}$}
\newcommand{\sys}{CertiBlock}

\theoremstyle{definition}
\newtheorem{definition}{Definition}
\theoremstyle{plain}
\newtheorem{theorem}{Theorem}

\lstdefinestyle{scr}{basicstyle=\ttfamily\footnotesize,columns=fullflexible,
  keepspaces=true,breaklines=true,xleftmargin=1.2em,frame=none,aboveskip=4pt,belowskip=2pt}

\begin{document}

\title{Certificate-Carrying Transformation of Event-Driven\\Block Programs}

\author{
\IEEEauthorblockN{Yuan Si and Jialu Zhang$^{*}$\thanks{Corresponding author: Jialu Zhang.}}\\
\IEEEauthorblockA{University of Waterloo, Waterloo, Canada\\
\texttt{yuan.si@uwaterloo.ca}, \texttt{jialu.zhang@uwaterloo.ca}}
}

\maketitle

\begin{abstract}
Block-based end-user languages such as Scratch run tens of millions of programs whose
authors are not expected to read a correctness argument. The tools that rewrite these programs
establish behavior preservation through program analysis and testing, without a checked guarantee. We turn optimization of such programs into
certificate-carrying source-to-source rewriting. An untrusted optimizer proposes a rewrite;
a trusted, fail-closed checker accepts it only after recomputing every side condition that the
rewrite's behavior preservation depends on, under an explicit observation lens. The checker
is the sole authority: given a correct checker and a small, explicitly stated set of model-to-VM
assumptions, an optimizer bug cannot mint an unsound acceptance. The
observation lens is a parameter of the framework, and the central soundness argument is a
cooperative-frame refinement theorem: a write whose value is overwritten before any thread
observes it, within a window in which no thread yields, can be removed. We mechanize this theorem in
Lean and show that one parametric statement covers two concrete rewrite families, instantiated to
variable state and to renderer state. We build a checker for six rewrite
families and evaluate it on 300 real Scratch projects sampled from the public repository. The
checker accepts a behavior-preserving rewrite on 94.3\% of projects (283 of 300), a breadth carried by
editor and asset cleanup with the certified-runtime families applying on fewer; certification
costs under one tenth of a second per project on average; and a cross-family adversarial campaign of
4{,}278 perturbed rewrites produces zero false accepts on the hardened checker. An independent
pre-submission audit had earlier found eight false accepts the per-family test suites missed; each is
now rejected and carries a regression test. An ablation that strips the semantic side conditions,
leaving the analysis-and-testing stance alone, ships rewrites that the virtual machine confirms
change behavior; the full checker rejects every one. The result is a working account of how to give
behavior-preservation guarantees for a concurrent, event-driven, end-user language.
\end{abstract}

\section{Introduction}

A Scratch project is a concurrent, event-driven program. Scripts run under a cooperative
scheduler, communicate through broadcasts and clones, and render to a stage that the player
watches frame by frame. The public Scratch repository holds tens of millions of such
projects, written largely by children and hobbyists, an audience not expected to inspect a
static analysis or a correctness argument. Programs in this setting carry the same kinds of redundancy that
compilers remove from conventional code: a sprite is repositioned twice with no visible
frame in between, a broadcast reaches no receiver, a costume is removed yet still listed.
A tool that cleaned up these programs would benefit authors who lack the means to do so themselves;
recurring quality problems in shared Scratch code are documented to impede its reuse and
sharing~\cite{techapalokul2017quality,robles2017quality}. The obstacle is trust: an optimizer that
silently changes a child's game is worse than none, and this population cannot notice or recover from
the change.

End-user languages make this trust problem sharp.
A professional developer reads diffs, runs tests, and reverts a bad commit; the author of a
Scratch game has none of these defenses, and the program is itself the deployed program, so a
source rewrite is the shipped change. A guarantee that a rewrite preserves what the player sees is
the only acceptable basis for touching these programs automatically, and producing one for a
concurrent, rendered, event-driven language is this paper's problem.

Certified optimization addresses this kind of trust gap, and it has a mature home in verified
compilers. CompCert proves once and for all that its passes preserve the observable behavior
of sequential C~\cite{leroy2009compcert,leroy2009backend}, and translation validation discharges
the same guarantee per run by checking each output against its
input~\cite{pnueli1998tv,necula2000tv,lopes2021alive2}. These techniques target sequential
intermediate representations inside a compiler. The programs we care about are cooperatively
concurrent, event-driven, and edited directly by their authors, and their observable behavior
is what a player sees and hears rather than a sequence of I/O calls. Certified rewriting for a concurrent end-user
language is open territory, and the legality of removing a write depends on the scheduler, clone
creation, what the renderer commits to the screen, and which editor state a tool may change. Automated refactoring for Scratch already exists and establishes behavior
preservation through program analysis, preconditions, and
testing~\cite{techapalokul2019refactoring,adler2021readability}. \sys{} adds a checker that
recomputes each side condition and is the sole minter of acceptances, under an explicit lens.

We present \sys{}, a system that rewrites Scratch programs under a trusted checker that certifies
each rewrite behavior-preserving relative to an explicit observation lens. The design follows one discipline: the optimizer proposes and the checker
disposes. The optimizer is free to be large, heuristic, and wrong. It emits a certificate that
names the rewrite family, the observation lens, and the concrete edit. The checker re-parses
the before and after programs, recomputes from scratch every side condition the family's
soundness depends on, and mints an acceptance only when all of them hold. Nothing the optimizer
claims is trusted, including its own digests. The soundness of the whole system then rests on an
auditable checker and two named model-to-VM conformance assumptions, while the large optimizer stays
untrusted and can evolve without enlarging the trusted base. The two core decision procedures, the
structural single-allowed-delta obligation and the cooperative-frame dead-store admissibility, are
realized as compiled Lean functions proved sound, a verified-executable reference for the deployed
Python checker, which the adversarial campaign, the audit, and the oracle validate against the model.

Two design choices carry the framework. The first is that behavior preservation is stated
relative to an explicit observation lens, and the lens is a parameter. A player observes
rendered frames, sounds, prompts, visible monitors, and termination; an editor additionally
observes comments and layout. A rewrite that is sound for a player is unsound for an editor,
and the checker enforces the distinction by pinning each family to the lens under which its
edit is behavior-preserving. The second is a cooperative-frame refinement theorem, which we
mechanize in Lean. The theorem isolates the property that makes dead-write removal sound under
cooperative scheduling: within a window in which no thread yields and no observation of the
written state occurs, an overwritten write is invisible. We state the theorem so that the
observed resource and the boundary at which observation happens are both parameters, and we
instantiate it to variable state observed at yield boundaries and to renderer state observed
at frame boundaries.

We implement six rewrite families spanning editor cleanup, event-graph simplification, and two
dead-store families, together with an equality-saturation optimizer that the checker gates
entirely and a differential oracle that runs the real Scratch virtual machine before and after
each accepted rewrite. We evaluate on 300 projects sampled uniformly from the public repository.
The study answers four questions: whether adversarial and VM-backed validation finds a false accept,
how often the checker accepts a behavior-preserving rewrite on real programs, how much it costs, and
how large the resulting trusted base is. A recurring finding shapes the evaluation. Per-family testing is not enough. An
independent audit of the integrated checker found eight false accepts that the per-family suites
had passed, each from a missed observation channel that only a cross-family adversary surfaced.

This paper makes the following contributions. We formulate optimization of event-driven end-user
programs as lens-parametric certificate-carrying rewriting, with a trusted checker that
recomputes side conditions and is the sole minter of acceptances (Section~\ref{sec:framework}).
We give a cooperative-frame refinement theorem, mechanized in Lean depending only on its kernel
axioms; the end-to-end claim rests separately on two named model-to-VM conformance assumptions. One
parametric statement covers variable and renderer dead-store elimination, and the two core decision
procedures are realized as compiled Lean functions proved sound against it, a verified-executable
reference for the deployed checker (Section~\ref{sec:theorem}). We build a checker for six families and an untrusted
optimizer gated by it (Section~\ref{sec:library}). We evaluate on 300 real projects and report
soundness, applicability, cost, and trusted-base size; a cross-family audit found eight false accepts
the per-family suites missed, with the transferable lesson that a multi-family checker needs a
boundary-crossing adversary (Sections~\ref{sec:impl}--\ref{sec:eval}).

\section{Background and Motivation}
\label{sec:background}

\subsection{Scratch as an event-driven language}

A Scratch program~\cite{maloney2010scratch,resnick2009scratch} is a set of sprites and a stage, each
carrying a graph of blocks. Execution begins at hat blocks such as ``when green flag clicked'' or
``when I receive m''. The runtime
runs all active scripts under a cooperative scheduler. A script executes straight through until
it reaches a yield point, at which control returns to the scheduler and the next frame may be
rendered. The yield points are a fixed set: the wait and wait-until blocks, every loop
iteration boundary, the glide blocks, asking for input, and the broadcast-and-wait block.
Ordinary motion and looks blocks do not yield, so a run of such blocks executes as one
uninterrupted segment.

A project is stored as a JSON document mapping block identifiers to blocks, where each block records
its opcode, inputs, fields, and the identifiers of its parent and next blocks. Two properties of this
representation shape the checker. Identity is by identifier: a variable, list, or broadcast message
carries a display name the author can change while the virtual machine binds by identifier, so a
check that compares display names can disagree with the machine. Some blocks appear in a compressed
array form, and a scan that does not expand them can miss a reference, so the checker rejects a
program whose blocks it cannot fully read.

Scripts interact through three mechanisms that the rewriting families must model. A broadcast
sends a message, which starts every receiver script for that message; the send does not
preempt the sender, and a plain broadcast returns immediately while broadcast-and-wait yields
until the receivers finish. A clone block copies the source sprite's full state into a new
instance and schedules the clone's start scripts, so the clone is born carrying the sprite's
position, costume, size, direction, visibility, and effects at the moment of creation. The
stage is redrawn once per scheduler step from the current sprite state, and the player observes
the program through that sequence of frames together with sounds, answer prompts, and the
values of visible variable and list monitors.

This execution model is the source of both the opportunity and the difficulty. The opportunity
is that the cooperative scheduler creates windows of straight-line execution in which
intermediate state never reaches the screen, so a write that is overwritten inside such a window
is invisible. The difficulty is that the boundaries of these windows are subtle and the
observation channels are many.

The observation channels divide into two kinds, and a sound checker accounts for both. The first is
the frame boundary: the renderer draws after every scheduler step, not only at program yields, so the
window in which no frame is drawn is exactly one uninterrupted segment, and any write whose overwriter
lies past a scheduler step is observable at the intervening draw. The second is the in-segment commit,
where an action makes state observable without yielding: a pen-down motion draws a line, a stamp
copies the sprite onto the stage, a clone snapshots the sprite, and a sensing block or a say bubble
reads or renders the sprite at the moment it runs. A yield ends a segment; an in-segment commit does
not, yet a yield-only analysis misses it. That distinction is the technical crux of the
visual dead-store family. Table~\ref{tab:channels} lists the channels and how a rewrite must treat
each, and the rest of the paper recomputes exactly these conditions.

\begin{table}[t]
\centering
\caption{Observation channels of the cooperative, rendered model and the
treatment a sound rewrite owes each.}
\label{tab:channels}
\begin{tabular}{@{}p{0.30\columnwidth}p{0.27\columnwidth}p{0.31\columnwidth}@{}}
\toprule
Channel & Modeled as & Treatment \\
\midrule
frame draw, per scheduler step & observation boundary & a visual write cannot cross it \\
pen down, stamp & in-segment commit & visual store in the window is live \\
clone creation & state snapshot & visual store in the window is live \\
visible variable or list monitor & player observation & the variable or list is live; asset removal rejects \\
costume or backdrop index read & player observation & asset removal that shifts it rejects \\
timer, mouse, loudness sensing & external read & store in the window is live; oracle fixes them with a deterministic harness \\
broadcast with no receiver & no player effect & dead-broadcast removal accepts \\
\bottomrule
\end{tabular}
\end{table}

\subsection{A motivating example}

Consider a sprite whose green-flag script places itself, sets up its costume, then places itself
again before entering its main loop.

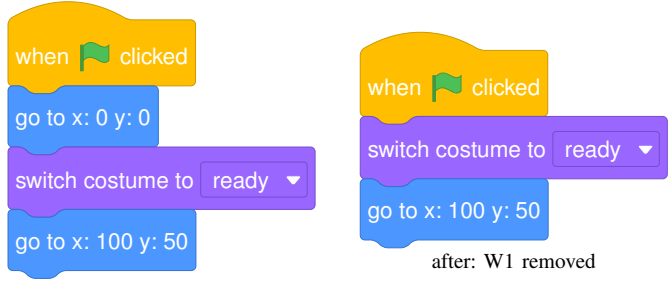
\begin{center}
\vspace{-4pt}
\begin{minipage}{0.46\columnwidth}\centering
\resizebox{\linewidth}{!}{%
\begin{scratch}
\blockinit{when \greenflag clicked}
\blockmove{go to x: 0 y: 0}
\blocklook{switch costume to \selectmenu{ready}}
\blockmove{go to x: 100 y: 50}
\end{scratch}}\\[2pt]
{\footnotesize before: W1 then W2}
\end{minipage}\hfill
\begin{minipage}{0.46\columnwidth}\centering
\resizebox{\linewidth}{!}{%
\begin{scratch}
\blockinit{when \greenflag clicked}
\blocklook{switch costume to \selectmenu{ready}}
\blockmove{go to x: 100 y: 50}
\end{scratch}}\\[2pt]
{\footnotesize after: W1 removed}
\end{minipage}
\vspace{-4pt}
\end{center}

The first placement W1 (\texttt{go to x:\,0 y:\,0}) is overwritten by W2 before the loop and no block
between them yields, so no frame is drawn at the origin and removing W1 preserves what the player
observes. The removal is sound only under the channels of Table~\ref{tab:channels}: a pen line or a
clone between W1 and W2 would make the intermediate position visible, a sensing read would leak it to a
monitor or a bubble, and a stage-edge clamp depending on the prior position would make the two
placements non-interchangeable. A checker that accepts the removal must recompute all of these, and a
theorem must establish they suffice. The rest of the paper builds that checker and that theorem.

\section{Lens-Parametric Certified Rewriting}
\label{sec:framework}

\begin{figure}[t]
\centering
\begin{tikzpicture}[
  font=\footnotesize,
  box/.style={draw, rounded corners=2pt, align=center, inner sep=4pt, minimum height=7mm},
  trust/.style={box, fill=black!6},
  untrust/.style={box, fill=white},
  >={Stealth[length=2mm]}]
\node[untrust] (prog) {program};
\node[untrust, right=5mm of prog] (opt) {optimizer\\\textit{(untrusted)}};
\node[trust, right=6mm of opt] (chk) {checker\\\textit{(trusted)}};
\node[trust, right=6mm of chk] (rec) {receipt\\\textsc{accept}/\textsc{reject}};
\node[untrust, below=5mm of chk] (vm) {VM oracle\\\textit{(untrusted)}};
\draw[->] (prog) -- (opt);
\draw[->] (opt) -- node[above,font=\scriptsize]{cert.} (chk);
\draw[->] (chk) -- (rec);
\draw[->] (chk) -- (vm);
\draw[->] (vm.west) to[bend left=18] node[below,font=\scriptsize]{conformance} (prog.south);
\end{tikzpicture}
\caption{The optimizer proposes a rewrite with a certificate; the checker recomputes every side
condition and mints the only authoritative verdict; the VM oracle independently checks behavior on
the real virtual machine. Shaded boxes are the trusted base.}
\label{fig:pipeline}
\end{figure}
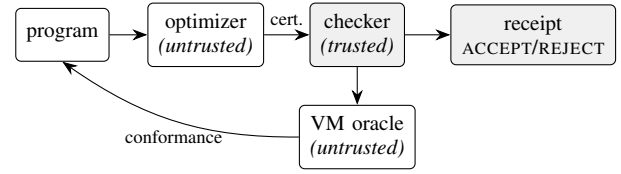

\subsection{The trusted boundary}

\sys{} separates an untrusted optimizer from a trusted checker, as Figure~\ref{fig:pipeline} shows.
The optimizer reads a program, proposes a rewrite, and emits a certificate. A certificate names the rewrite family, the
observation lens, a locator for the edited region, and the SHA-256 digests of the before and
after byte streams; the certificate is a set of hints, a locator and not a proof object. The checker re-parses both byte
streams, recomputes the digests, pins the lens from its name through a fixed table, looks up the
family's obligations in a registry, and discharges each obligation against the recomputed facts.
The checker mints a receipt with a single authoritative verdict, and only a receipt that records
acceptance permits the rewrite to ship. This is the LCF discipline applied to rewriting. The
checker is the only component that can declare a rewrite sound, and it does so by reconstructing the
fact, never by accepting the optimizer's word for it.

The receipt the checker mints carries the verdict and the obligation results, and no field of a
certificate can set a verdict. A family is dischargeable only if it is registered in the checker, and
its obligation list is fixed by the family author and read from the registry. The optimizer cannot
add, remove, or reorder obligations, and an unregistered family name is rejected outright. Each family
carries one obligation of a distinguished kind, the single-allowed-delta obligation, which proves that
the only structural difference between the before and after programs is the declared edit class. Any
change the optimizer introduces beyond the declared edit fails it, so the checker need not anticipate
every way an optimizer might misbehave; it only confirms that the difference it sees is the one the
family is allowed to make. The checker also wraps every obligation in a fail-closed guard that turns
an exception into a rejection, so the trusted base degrades safely.

Algorithm~\ref{alg:check} states the entire trusted procedure. It re-derives the digests from the
byte streams before reading anything else, so a certificate that misreports the program it edited
is rejected at once. It pins the lens by name and looks the family up in the registry, so an unknown
lens or an unregistered family is rejected before any obligation runs. It then discharges the
family's fixed obligation list and accepts only when every obligation proves. The procedure is the
sole minter of an acceptance, and every branch that is not an explicit acceptance is a rejection,
which is what makes the trusted base fail closed.

\begin{algorithm}[t]
\DontPrintSemicolon
\caption{Trusted check of a single rewrite}
\label{alg:check}
\KwIn{certificate $c$; before and after bytes $b, a$}
\KwOut{a single verdict, \textsc{Accept} or \textsc{Reject}}
\lIf{$\mathrm{sha}(b)\neq c.bd$ \textnormal{ or } $\mathrm{sha}(a)\neq c.ad$}{\Return \textsc{Reject}}
$P \leftarrow \textsc{Parse}(b)$;\quad $P' \leftarrow \textsc{Parse}(a)$\;
$L \leftarrow \textsc{Lens}(c.\mathit{lens})$;\quad $F \leftarrow \textsc{Registry}[c.\mathit{family}]$\;
\lIf{$L$ \textnormal{ or } $F$ \textnormal{ undefined, or } $F.\mathit{lens}\neq L$}{\Return \textsc{Reject}}
\ForEach{obligation $o \in F.\textsc{Obligations}(L)$}{
  \lIf{$\textsc{Discharge}(o, P, P', L)\neq \mathtt{proved}$}{\Return \textsc{Reject}}
}
\Return \textsc{Accept}\;
\end{algorithm}

\subsection{Observation lenses}

Behavior preservation is meaningful only relative to what counts as behavior. \sys{} fixes a
single raw observation of an execution and defines lenses as views of it. The player lens
\lplayer{} observes the sequence of rendered frames, which the runtime commits at every scheduler
step, together with sounds, answer prompts, the values of visible variable and list monitors, and
termination. Broadcasts and clones are observable only through these channels, so a broadcast that
reaches no receiver, and produces no frame, sound, prompt, or monitor change, is
unobservable under \lplayer{}. The final-state lens \lfinal{} observes selected final state under
bounded scenarios and ignores intermediate frames. The debug lens \ldebug{} additionally observes
editor and comment state. The lenses form a tower in which
\ldebug{} is finer than \lplayer{}, which is finer than \lfinal{}: an observation distinguished by
a coarser lens is distinguished by every finer lens. Refinement of one program by another is
inclusion of observable trace sets under a lens, so a rewrite preserves behavior under a lens when
the after program refines the before program at that lens.

The framework takes the lens as a parameter, which lets it host families with different soundness
conditions under one mechanism. Editor cleanup is sound under \lplayer{} because
removing a comment leaves the canonical program core byte-identical, and it is unsound under
\ldebug{} because an editor observes comments. The checker pins comment cleanup to \lplayer{} by
name, so a certificate claiming any other lens is rejected before obligations run. A dead-store
family is also pinned to \lplayer{} but for a different reason. Its edit changes runtime state
that a player would observe if the store were live, and the family's obligations prove that the
store is not live. Size is a property a tool reports as a metric, and it is not a lens; the
framework measures byte savings outside the behavioral model. The result is a uniform treatment:
each family declares the lens under which its edit is behavior-preserving, and the checker holds
it to that declaration.

\subsection{Side-condition recomputation}

The defining move of the framework is that the checker recomputes side conditions and trusts nothing
in the certificate. Recomputation has a concrete payoff in this domain, because the
soundness of a rewrite depends on how the Scratch virtual machine reads the program, and the
checker must read it the same way. Three examples show why a plausible check is wrong and a
recomputed one is right.

Broadcast dispatch resolves a sent message by identifier through the project's broadcast table
and matches messages case-insensitively. A message that the author renames in the editor keeps
its identifier while its visible name changes, so a reachability check that compares visible
names concludes that a live send is dead. The checker resolves the send by identifier through the
table, the way the virtual machine does, and matches case-insensitively over the ASCII range
where the comparison is provably the same as the machine's. A message name outside that range
makes the case folding uncertain, and the family rejects.

The sensing block that reads another sprite's attribute takes its target sprite from an input, and
a check that reads the block's field instead misidentifies the observed sprite, so it can miss a
read of a variable it is about to declare dead. The checker reads the target from the
input the virtual machine consults. The third example is asset removal. Removing a costume below
the current index shifts the one-based index that another script reads through a sensing block or
a visible monitor, so the family treats any such reader as an observer of the index and rejects a
removal that would shift it. Each of these is a place where the checker reads exactly what the
machine reads, and where it cannot, the responsible family rejects. This default-deny stance is
the reason the trusted base can be small: the checker never approximates a machine operation it
cannot replicate.

The same stance governs unsupported features. The Scratch opcode space is open, so the checker
classifies blocks by an allow-list of known stock opcodes, and any extension or unknown opcode falls
outside it and forces the responsible family to reject. A cloud variable is treated as always live,
a read of timer, mouse, loudness, or current time counts as an observation of external state that
keeps a write in its window live, and a computed broadcast name refuses the broadcast-reachability
families. An unsupported feature is
rejected or marked unknown, never silently assumed safe, so the soundness of the checker does not
depend on having enumerated every feature of an evolving platform.

\section{The Cooperative-Frame Theorem}
\label{sec:theorem}

\subsection{Statement}

The technical core of \sys{} is a refinement theorem for dead-store elimination under cooperative
scheduling. The theorem isolates the condition under which an overwritten write is unobservable.
Consider a thread that writes a resource, continues without yielding, and writes the same resource
again. Call the span between the two writes the pending window, the first write the dead store, and
the second the overwriter. If no observation of the resource occurs in the pending window and the
window contains no yield, then the state the dead store establishes never reaches an observation,
and deleting the dead store refines the original program under \lplayer{}.

The theorem is stated with two parameters: the observed resource and the boundary at which
observation happens. A boundary is a step that reads a set of resource cells and emits a trace
event, and the same abstraction covers a read by the running thread and a frame drawn by the renderer.

\begin{definition}[Pending window and admissibility]
Let $W_1$ write resource cell $r$ and let $W_2$ be the next write of $r$ on the same thread with
no yield between them. The pending window is the span of steps between $W_1$ and $W_2$. The window
is admissible when every observation step $o$ in it satisfies $\mathrm{read}(o) \cap \{r\} =
\emptyset$, and $W_2$ does not read the value of $r$.
\end{definition}

\begin{theorem}[Cooperative frame]
If the pending window of $W_1$ is admissible, then the program obtained by deleting $W_1$ refines
the original under \lplayer{}.
\end{theorem}

Cooperative scheduling makes the no-yield window single-threaded, so the only writes to $r$ within it
are $W_1$ and $W_2$, and a concurrently scheduled thread enters only at a yield, which ends the window.
The admissibility condition states that no observation in the window reads the dead cell and that
the overwriter does not depend on the old value. Under it the trace produced with the dead store
present equals the trace produced with it deleted. The proof exhibits a simulation between the two
programs in which the configurations agree on every cell except the dead one, and an admissible
observation, reading no dead cell, cannot distinguish the configurations, so the emitted traces
coincide step by step until the overwriter restores agreement on the dead cell as well. The window
need not be straight-line: an intervening conditional whose branches neither yield nor observe $r$
stays within the segment, and the mechanization certifies that case too.

\subsection{Two instances of one theorem}

The two dead-store families of \sys{} are instances of this single statement at different choices
of resource and boundary. Variable dead-store elimination instantiates the resource to a variable
cell and the boundary to a yield, because a player observes a variable only through a visible
monitor recomputed at frame boundaries. Visual dead-store elimination instantiates the resource to
a renderer cell, a sprite's position, direction, size, visibility, costume, or graphic effect, and
the boundary to a renderer draw. The two families are the same theorem applied at two boundary
values: the variable corollary and the visual corollary are the parametric theorem specialized to the
two boundaries, checked by reflexivity.

Parameterizing the boundary is what lets one proof serve both. A yield-only theorem is sound for
variables and unsound for renderer state, because renderer state commits at points a yield-only
analysis does not see. A new dead-store family supplies a resource and an observation set and
inherits the proof. We claim only the genericity these two instances exercise; a third resource would
test it further.

The renderer boundary forces the theorem's generality to do real work. The virtual machine draws a
frame after every scheduler step, so the only window without a draw is a single uninterrupted segment,
and the observation boundary for visual state is larger than a program yield. The in-segment commits
of Table~\ref{tab:channels}, a pen line, a stamp, or a clone creation, each count as an observation in
the pending window, so a dead visual write whose window contains one fails admissibility and the
family rejects it. The model also requires the overwriter to be commit-pure (it produces no in-segment commit and ignores
the old value), which matters for motion:
the virtual machine clamps a placement against the stage edge, and the clamped result can depend on
the prior position, so the family admits a motion overwrite only when its coordinates are literals
within the stage bounds, where the clamp is the identity. We verify in Lean that a pending-window
observation of the dead resource makes the removal unsound and that admissibility rejects exactly
that case.

\subsection{Mechanization and trusted base}

The development is mechanized in Lean~4 (v4.30.0)~\cite{moura2021lean4} without external libraries. The step
relation, the yield points, the observation steps, and the traces are defined there; the paper states
the theorem in prose and the mechanized development carries the formal definitions, so the result is a checked
theorem and not a prose claim. The headline theorem and both family instances depend only on the
propositional-extensionality and quotient-soundness axioms that the Lean kernel provides, with no use
of classical choice and no family-specific modeling assumption. Editor-cleanup soundness is a theorem rather than an
axiom: the model defines a canonical core as the projection the virtual machine loads, discarding
comments, layout, and metadata, and the proof tracks the loader's read set to show that programs with
equal cores produce equal player observations, so a cleanup that changes only material outside the
core refines under the player lens, modulo the same loader-conformance assumption the dead-store
families rely on. Two model-to-VM conformance assumptions remain, and we name them as an explicit
trusted base, in the style of the axiomatized hardware of CompCert. The first states that a player
observation depends only on the projected runtime state, and the second states that the pinned virtual
machine loads a program in conformance with the model's loader. Both concern the fidelity of the model to
the real machine, and the differential oracle of Section~\ref{sec:eval} is the empirical evidence
for them; neither is proved. The Lean development is 5{,}909 lines and builds with no
incomplete proofs.

The mechanized theorem certifies an abstract model. The two core decision procedures are realized as
compiled Lean functions proved sound, a verified-executable reference. The deployed checker is a
separate Python implementation, connected to the model by construction and testing, and the
differential oracle tests the model's observation boundary against the machine. A verified bridge from
the reference to the deployed checker is future work, the honest limit of the guarantee treated in
Section~\ref{sec:threats}.

\section{The Rewrite Library}
\label{sec:library}

\subsection{Six families}

\sys{} implements six rewrite families that span editor cleanup, event-graph simplification, and
dead-store elimination. Comment cleanup removes comment entries and their block backreferences. It
is sound under \lplayer{} because the canonical program core, the projection that the virtual
machine loads, is unchanged, and the family proves core equality directly. Unused-asset cleanup
removes costumes, sounds, variables, and lists that no block references and de-duplicates
byte-identical assets. Its delicate condition is the costume index: removing an unreferenced costume
below the current one shifts the index that a sensing block or a visible monitor reads, so the
family forces a project-wide rejection of costume and backdrop removal whenever any such reader is
present.

Dead-broadcast removal deletes a broadcast send whose message reaches no receiver. The family
resolves the message by identifier through the broadcast table, matches receivers case-insensitively
over the ASCII range, and rejects when a computed message name or a non-ASCII name puts the reachable
set beyond what the checker can decide. Dead-code elimination removes whole unrunnable top-level
components: an empty handler hat on a fixed whitelist of stock trigger opcodes, a receiver hat for a
message that is never sent, or a loose stack with no triggering hat. The family computes the
component as the transitive closure of a head over its successor chain and its nested inputs, and it
refuses to cross into an independent top-level script, because the virtual machine runs any top-level
block as its own script regardless of an incoming successor edge.

The two dead-store families realize the cooperative-frame theorem. Variable dead-store elimination
removes a variable write that an overwriter cancels inside a no-yield window with no intervening
read, no external observation, and no cloud variable or visible monitor on the variable. Visual
dead-store elimination removes an overwritten renderer write under the conditions
Section~\ref{sec:theorem} requires: a single uninterrupted segment, an absolute overwriter of the
same key, no pen, stamp, clone, sensing, or bubble observation in the window, and for motion the
literal-in-bounds condition under which the stage clamp is the identity. All six are sound under the
player lens. Table~\ref{tab:families} states, per family, the condition the checker recomputes, whether
the soundness rests on a Lean-mechanized theorem, and the oracle's role. Three of the six carry a
mechanized theorem (cooperative-frame for the two dead-store families, core-equality for comment
cleanup); the other three rest on a recomputed structural argument the oracle corroborates.

\begin{table}[t]
\centering
\caption{The six families: the condition the checker recomputes, whether the soundness rests on a
Lean-mechanized theorem, and the differential oracle's role (sanity, seeded, or VM validation).
All are sound under the player lens.}
\label{tab:families}
\begin{tabular}{@{}p{0.30\columnwidth}p{0.36\columnwidth}cl@{}}
\toprule
Family & Recomputed condition & Mech. & Oracle \\
\midrule
Comment cleanup & canonical core unchanged & yes & sanity \\
Unused-asset cleanup & no reference or index-shift observer; de-dup retargets references; clean splice & no & sanity \\
Dead-broadcast removal & message reaches no receiver & no & sanity \\
Dead-code elimination & component never runs; clean splice & no & sanity \\
Variable dead-store & overwritten in a no-yield window & yes & seeded \\
Visual dead-store & no observable commit in the segment & yes & VM \\
\bottomrule
\end{tabular}
\end{table}

\subsection{The visual dead-store decision procedure}

The visual family turns the cooperative-frame theorem into a decision procedure, stated as
Algorithm~\ref{alg:vds}, whose steps map one to one onto the clauses of the theorem. The procedure
walks the straight-line segment after the candidate write and accepts only when the segment realizes
the admissibility condition. The pen guard removes the one in-segment commit the walk cannot localize,
because a pen-down state makes every motion an observable commit. The window-breaker test ends the
segment at the first yield, where a frame is drawn; the observer test is admissibility itself, refusing
the removal when a sensing read, stamp, clone, or bubble in the window reads the key or a coupled key.
The overwriter must be a full absolute commit-pure write, the theorem's requirement that it not read
the old value, and a partial overwrite is refused. The fence-identity test enforces the one VM-level
subtlety the theorem leaves to the checker: a motion overwrite is a true overwrite only when the stage
clamp is the identity, which holds for literal in-bounds coordinates on a sprite that cannot be
dragged. The clean-splice test confirms the proposed program $P'$ is exactly $P$ with the candidate
removed.

\begin{algorithm}[t]
\DontPrintSemicolon
\caption{Visual dead-store acceptance}
\label{alg:vds}
\KwIn{program $P$, proposed rewrite $P'$; candidate write $W_1$ on renderer keys $K$, target $t$}
\KwOut{\textsc{Accept} if $W_1$ is a removable dead store, else \textsc{Reject}}
\lIf{\textsc{PenPresent}($P$) \textnormal{ and } \textsc{IsMotion}($W_1$)}{\Return \textsc{Reject}}
$W_2 \leftarrow \bot$\;
\ForEach{$x \in \textsc{StraightLine}(W_1)$}{
  \lIf{\textsc{Breaker}($x$) \textnormal{ or } \textsc{Observes}($x, K$)}{\Return \textsc{Reject}}
  \uIf{\textsc{FullOverwrite}($x, K$)}{$W_2 \leftarrow x$; \textbf{break}}
  \lElseIf{\textsc{Writes}($x, K$)}{\Return \textsc{Reject}}
}
\lIf{$W_2 = \bot$}{\Return \textsc{Reject}}
\lIf{\textsc{IsMotion}($W_2$) \textnormal{ and not } \textsc{FenceId}($W_2, t$)}{\Return \textsc{Reject}}
\lIf{\textsc{Core}($P'$) $\neq$ \textsc{Core}(\textsc{Remove}($P, W_1$))}{\Return \textsc{Reject}}
\Return \textsc{Accept}\;
\end{algorithm}

\subsection{Obligations and structural soundness}

Algorithm~\ref{alg:vds} is one instance of a decision-procedure skeleton the families share, and the
others differ only in the segment they walk and the observation they forbid. Variable dead-store
elimination walks the same straight-line segment but forbids a read of the variable rather than a
renderer observation, and adds the cloud-variable and visible-monitor checks that make the variable
externally observable. Dead-broadcast and dead-code removal walk the event graph, and their forbidden
observation is a reachable receiver or trigger. This shared skeleton is what lets the cooperative-frame
theorem serve more than one family: each names the segment, the resource, and the forbidden
observation, and inherits the rest.

Each family is a list of obligations the checker discharges, and the lists share a shape.
A lens check confirms that the claimed lens is the one the family is sound under. A well-formedness
pre-pass
rejects compressed or malformed block forms that the scans cannot see, because a block the scan
cannot read could hide a reference, a sender, or a triggering hat. The single-allowed-delta
obligation confines the structural difference to the declared edit. The family-specific conditions
then establish behavior preservation under the lens.

Structural removal is reconstructed rather than trusted. The checker derives the set of removed block
identifiers from the difference between the before and after programs, recomputes the program that
results from deleting exactly that set, and requires its canonical core to equal the after program's
core. A partial deletion, an orphaned successor, or a dangling reference changes the reconstructed
core and fails the comparison, so the checker accepts a removal only when the after program is
exactly the before program with a clean splice of the declared component. This reconstruction is the
same for every removal family, which is why the soundness argument for structural edits is shared
rather than re-proved per family.

\subsection{The optimizer and the differential oracle}

The optimizer is an equality-saturation engine~\cite{tate2009egraph,willsey2021egg} that proposes
rewrites and keeps only those the checker accepts. Saturation gates every proposed edge through the
checker, so a hostile gate that rejects everything yields the unchanged input, and the composed
result is re-checked as a sequence against the actual intermediate programs. The cost model lives in
the optimizer and chooses among already-accepted states, so it cannot make an unsound result win.
The optimizer is outside the trusted base, and it plays no role in the soundness argument; the edits
themselves are framed graph rewrites in the algebraic graph-transformation
tradition~\cite{ehrig2006dpo}.

Composing several rewrites needs care because a certificate justifies one step. The optimizer emits
a sequence certificate with the digests of the intermediate programs, and the checker re-anchors on
the actual intermediates and checks each step against the recomputed program the previous step
produced. Composition stays within one lens, and the unification of the families on the player lens
lets one saturation run compose a cleanup rewrite with a dead-store rewrite into a single checkable
sequence: each step is certified independently against the recomputed intermediate, and the
transitivity of the refinement preorder, mechanized in Lean, lifts the per-step guarantees to the
whole sequence. This composition is where the engine earns its place over a single-pass applier: on 57
of the 300 projects the verified result composes two or more families, up to three families and seven
steps in one sequence, which a one-rewrite-per-family pass could not express.

Behavior preservation of an accepted rewrite is checked a second way by a differential oracle that
runs the real Scratch virtual machine on the before and after programs under identical deterministic
conditions and compares the player-visible traces. The oracle is a sanity check on the model and is
itself untrusted. Two of its properties matter for the evaluation. It observes only player-visible
state, so it ignores editor bookkeeping such as an unreferenced broadcast-message variable that the
editor garbage-collects on load. It runs a determinism control on any flagged mismatch, re-tracing
the before program several times, because a small fraction of real projects combine randomness,
waits, and clones in a way the headless harness cannot reproduce deterministically; a project whose
own behavior is not reproducible is reported as not adjudicable rather than as a mismatch. The oracle
catches discrepancies between the model and the machine, and the audit of Section~\ref{sec:eval}
shows that this second method finds mistakes the static obligations alone do not.

\section{Implementation and Trusted Base}
\label{sec:impl}

\sys{} implements the checker and optimizer in Python, drives a headless build of the Scratch
virtual machine for the differential oracle, and mechanizes the proofs in Lean~4. The trusted base
is the checker package, which the soundness claim rests on. It is 9{,}921 lines and divides into three
parts: a fixed kernel of 1{,}856 lines for the certificate and receipt types, the lens table, the
resource model, the well-formedness pre-pass, and the family registry; 6{,}060 lines of family
obligations across the six families; and a 2{,}005-line intermediate-representation compiler that only
the variable dead-store family needs to reason about variable reads and writes across control flow.
The untrusted optimizer (3{,}060 lines) and the differential oracle and harness (2{,}292 lines)
together are 5{,}352 lines and carry no part of the guarantee. The two core decision procedures are also
realized as compiled Lean functions proved sound (Section~\ref{sec:theorem}), a verified-executable
reference separate from these 9{,}921 deployed Python lines.

The proportion is the point of the design. A bug anywhere in the 5{,}352 untrusted lines, or in the
optimizer's heuristics and cost model, cannot cause an unsound acceptance, because the checker
re-derives every fact it relies on, so the optimizer can grow toward more aggressive rewriting without
enlarging what a reviewer must trust. The family obligations dominate the trusted base because each
encodes the observation conditions of its edit class, where the domain knowledge of the cooperative,
rendered model lives. The parser is trusted and fails loudly on any input it cannot represent, and the
obligation runner wraps every obligation in a fail-closed guard, so a malformed program or a defect in
one family degrades to a refusal rather than an unsound acceptance. When the checker is unsure it
rejects, and the cost of that discipline is the completeness reported as the prevalence numbers.

\section{Evaluation}
\label{sec:eval}

We evaluate \sys{} on 300 projects sampled from the public Scratch repository. We draw project
identifiers uniformly at random across the version-3 id range. Most identifiers in the range resolve
to deleted, unshared, or empty projects, which we skip; we keep the shared, loadable ones until 300
are collected. This sampling avoids the popularity bias of the front-page listing, and the 300 parse
and load by construction. The corpus spans real projects, from single-sprite animations to games with
more than two hundred targets and over fourteen thousand blocks; the median project has 3 targets and
34 blocks, the 90th percentile 556 blocks. Across the sample 35\% use broadcasts, 19\% clones, 17\%
custom blocks, and 9\% pen, so it exercises the concurrency, clone, and rendering features the checker
reasons about. We ask four questions. RQ1 asks whether the
checker admits a false accept. RQ2 asks how often the checker accepts a behavior-preserving rewrite on
real programs. RQ3 asks what certification costs. RQ4 asks how the formal development bounds the
trusted base.

\subsection{RQ1: Does the checker admit a false accept?}

The Lean theorem establishes soundness for the abstract model; RQ1 asks whether the executable checker
and the model-to-VM boundary admit a false accept under adversarial and VM-backed validation. We
attack the implementation five ways: unit obligations per family, an independent audit of the
integrated checker, a cross-family adversarial campaign, VM-differential conformance, and an ablation
that removes the semantic side conditions to confirm the checker rejects the unsoundness they guard.

Each family carries a unit suite that exercises its accept path together with one adversarial
negative per obligation. The six suites hold 274 tests, and all pass. Passing the per-family suites
is necessary, and the audit shows it is not sufficient.

\subsubsection{The audit}

We ran an independent adversarial audit of the integrated checker. Each probe attacks one family
along one pitfall class drawn from the observation channels, and each finding is reproduced and
confirmed against the real virtual machine before it is counted. The audit found eight false accepts
in families that had passed their own suites (Table~\ref{tab:audit}), all on observation channels a
single-family view does not see, which is why the audit finds what the unit suites miss.

\begin{table}[t]
\centering
\caption{False accepts found by the audit, grouped by the observation channel
the family failed to recompute.}
\label{tab:audit}
\begin{tabular}{@{}llc@{}}
\toprule
Family & Missed observation channel & Count \\
\midrule
Unused-asset cleanup & costume/backdrop index read & 4 \\
Dead-code elimination & closure reach; broadcast timing & 2 \\
Visual dead-store & relative selector; yielding sound & 2 \\
\bottomrule
\end{tabular}
\end{table}

One false accept illustrates the class. Unused-asset cleanup removed an unreferenced costume below the
current index and re-pointed to keep the same image, but the removal shifted the one-based index of
every costume above it, which a second sprite read through a sensing block and displayed: the image was
unchanged while the number changed. The per-family suite tested removal and re-pointing, not a second
sprite reading the index, the channel that appears only when two families' concerns meet. The fix
treats any reader of the index, a sensing read, a monitor, or an extension, as an observer.

The other false accepts have the same shape, each a missed observation channel and each now rejected
with a regression test that keeps it closed.

\subsubsection{The adversarial campaign}

The cross-family adversarial campaign tests the single-allowed-delta backstop that every family
shares. For each project and family it takes the optimizer's proposed removal and derives variants
that add a second edit beyond the declared one, by flipping a literal, perturbing a field, inserting a
statement, or deleting an unrelated block; each variant is no longer the single allowed delta, so the
checker must reject it. Over the real 300-project corpus and a supplementary set of 400 real Scratch
projects from~\cite{scratchlens}, the campaign runs 4{,}278 such variants across the six families and
produces zero false accepts. The variable dead-store family, inert on natural projects, is
exercised by seeding a genuine dead store before perturbing it, and the bespoke comment and asset
fuzzers accept no invalid rewrite either. This campaign stresses the structural single-allowed-delta
backstop every family shares; the semantic side conditions of the dead-store families are exercised
separately by the differential oracle, so the two methods cover complementary halves of each acceptance.

\subsubsection{VM-differential conformance}

The differential oracle runs scratch-vm~1.6.19 on the before and after of every accepted rewrite,
driven from the green flag with a deterministic empty answer to any prompt, for 240 scheduler ticks
under a fixed seed, and compares the player-visible trace of frames, sounds, monitor values, and
termination. Table~\ref{tab:vmdiff} reports the result per family. Across the corpus the oracle finds no
behavioral mismatch on any accepted rewrite. Six accepted rewrites are reported as not adjudicable
because their own behavior is not reproducible under the headless harness, identified by the determinism
control; their guarantee rests on the static obligations, which hold without the oracle. The visual
family is evaluated under the full kinematic lens that records sprite position and costume, the variable
family on seeded dead stores because it is inert on natural projects, and the rest on their natural
acceptances.

\begin{table}[t]
\centering
\caption{VM-differential validation on the checker-accepted rewrites of the 300-project
corpus, all under \lplayer{} (visual under the finer kinematic projection, variable on seeded
mutants). Every accepted rewrite is adjudicated: validated plus flaky equals accepted, and no
accepted rewrite is a mismatch. ``Flaky'' is a project whose own behavior is not reproducible
under the harness.}
\label{tab:vmdiff}
\begin{tabular}{@{}lrrr@{}}
\toprule
Family & Accepted & Validated & Flaky \\
\midrule
Comment cleanup & 25 & 24 & 1 \\
Unused-asset cleanup & 267 & 264 & 3 \\
Dead-broadcast removal & 6 & 6 & 0 \\
Dead-code elimination & 56 & 55 & 1 \\
Variable dead-store (seeded) & 23 & 23 & 0 \\
Visual dead-store & 17 & 16 & 1 \\
\bottomrule
\end{tabular}
\end{table}

\subsubsection{The side conditions are load-bearing}

The first four methods show the checker accepts no unsound rewrite; this one shows its semantic side
conditions are why. We build a frame-unaware permissive remover that applies the textbook dead-store
rule to renderer writes, dropping the cooperative-frame model: it ignores intervening frame boundaries,
pen state, clone or stamp snapshots, and stage clamping, the analysis-and-testing stance of a tool
without a behavioral model. We run it on every project, take the real virtual machine as ground truth
for whether each removal changes the player-visible trace, and record the checker's verdict on the
unsound ones. On the 300-project corpus it fires 79 times; the oracle confirms 18 change behavior, and
the checker rejects all 18, in each case naming the violated side condition: a frame boundary
(\texttt{wait}, glide, a yielding sound), a clone snapshot, or an unbounded call. The result replicates on
400 supplementary projects from the ScratchLens dataset of student Scratch
projects~\cite{scratchlens} (177 removals, 33 unsound, all rejected).

The five methods answer RQ1 together: no method finds a surviving false accept, the audit now
carries the regression tests that keep its findings fixed, and the ablation shows the side conditions
catch real unsoundness, not merely assert safety. The cross-family audit was necessary to reach that state.

\subsection{RQ2: How often does the checker accept a rewrite?}

A guarantee is useful only if the checker accepts behavior-preserving rewrites on real programs.
The checker accepts at least one behavior-preserving rewrite on 283 of the 300 projects (94.3\%, 95\%
CI [91.1, 96.4]); the per-family counts in Tables~\ref{tab:vmdiff} and~\ref{tab:cost} are accepted
rewrites, and a project can take rewrites from several families. The untrusted optimizer's equality
saturation completes within the time bound on 286 of the 300 projects; the 14 that time out are a
scalability limit of the optimizer, not the checker, recoverable by a more scalable enumeration
without touching the trusted base.

The families differ in how often they apply, and the differences reflect the structure of real
Scratch. Editor cleanup and unused-asset cleanup are common, applying on most projects, because
authors leave comments and unreferenced assets behind as they iterate, and these cleanup families
carry the acceptance count. The runtime families apply less often and are where the certified-runtime
contribution lies. Among them, visual dead-store elimination has the most reach: the feasibility
scan finds candidates in 27 of the 300 projects (9.0\%, 95\% CI [6.3, 12.8]), spread across the visual
keys. The checker accepts a removal on 17 of them; the oracle validates 16 and reports 1 as not
adjudicable, with no mismatch. The result also generalizes to student code~\cite{scratchlens}: there,
dead stores to motion and graphic-effect keys are common, and every accepted removal is VM-validated
with no mismatch.
Dead-code and dead-broadcast removal apply on a smaller set of projects whose event graphs contain
unreachable handlers or sends. Variable dead-store elimination is inert on natural projects: a
variable in real Scratch is almost always read somewhere, so a transaction-local dead store is rare.

The corpus makes the inertness precise: 113 projects write a variable and 105 write one twice, yet none
yields a removable transaction-local dead store, the variable being read elsewhere first, so the
variable family is a proof-of-concept second instantiation of the theorem rather than a rewrite that
fires on natural code.

\subsection{RQ3: What does certification cost?}

Certification is cheap. Table~\ref{tab:cost} reports, per family, the kernel-accepted rewrites on the
corpus, the bytes saved on the accepted set, and the average checker time per project. Every family
mean is under one tenth of a second despite the checker recomputing every side condition; across
accepted checks the median is 2\,ms, the 95th percentile 66\,ms, and the slowest single check 1.3\,s
on the largest project. The
byte savings concentrate in the cleanup families because their edits remove larger structures; the
runtime families save little space because their value is behavioral. At this cost the checker can run
on every editor save, and as a background service across a repository.

\begin{table}[t]
\centering
\caption{Kernel-accepted rewrites, bytes saved, and average per-project checker time on the 300
projects, for the five families with natural acceptances (variable dead-store is inert;
see Table~\ref{tab:vmdiff}).}
\label{tab:cost}
\begin{tabular}{@{}lrrr@{}}
\toprule
Family & Accepts & Bytes saved & Time/proj. \\
\midrule
Comment cleanup & 25 & 36{,}626 & 77.0\,ms \\
Unused-asset cleanup & 267 & 365{,}857 & 6.3\,ms \\
Dead-broadcast removal & 6 & 1{,}389 & 28.0\,ms \\
Dead-code elimination & 56 & 246{,}159 & 51.2\,ms \\
Visual dead-store & 17 & 3{,}234 & 46.2\,ms \\
\bottomrule
\end{tabular}
\end{table}

\subsection{RQ4: How large is the trusted base?}

The trusted base is the 9{,}921-line checker of Section~\ref{sec:impl} together with the Lean
development and two model-to-VM conformance assumptions. The Lean development is 5{,}909 lines, builds
with no incomplete proofs, and its dead-store theorems depend only on the two kernel axioms that Lean
provides, with no classical choice and no family-specific modeling assumption. The two core decision
procedures are also realized as compiled Lean functions proved sound under Lean's standard
\texttt{propext}, \texttt{Quot.sound}, and the \texttt{decide}-introduced \texttt{Classical.choice}, a
verified-executable reference the deployed Python checker is validated against. The two conformance
assumptions are named statements about the model's fidelity to the virtual machine, with the RQ1 oracle
as empirical evidence. This is a CompCert-style trusted base:
an auditable checker, a mechanized argument, and a short list of named assumptions about the parts the
proof does not reach. Adding a family grows the trusted base by one obligation list audited in isolation,
so the trust cost grows linearly in the number of families.

\section{Discussion and Threats to Validity}
\label{sec:threats}

The deployed checker is a Python implementation of the model, so a transcription error there is a
soundness gap the proof does not close. The four RQ1 methods and the audit guard against it, and a
verified bridge from the Lean reference to the checker would close it.

The differential oracle is untrusted and can miss or over-report. For projects it cannot adjudicate,
the guarantee rests on the static obligations, the correct fallback: the oracle checks the model, not
the source of soundness.

The framework should carry to other event-driven run-to-completion languages, though we have not
instantiated a second substrate: the cooperative-frame theorem needs only a scheduler whose
observations commit at identifiable steps, not preemptive shared memory.

The corpus is a uniform random sample, so RQ2's prevalence numbers carry sampling uncertainty and, as
the families are conservative by design, lower-bound what a sound tool could accept. The player lens
approximates perceived behavior by frames, sounds, prompts, monitors, and termination, not exact pixels
or audio. We analyze only public project JSON and include no personal content.

\section{Related Work}
\label{sec:related}

\sys{} adopts the trusted-boundary discipline of verified compilation and the per-run stance of
translation validation, and applies them to a concurrent end-user language. CompCert proves a fixed
compiler correct once over small-step semantics with explicit event
traces~\cite{leroy2009compcert,leroy2009backend}, CakeML extends a single top-level correctness
theorem across a whole verified pipeline~\cite{kumar2014cakeml}, and later work carries verification
into separate compilation and the surrounding program logic~\cite{stewart2015compositional,appel2011vst}.
\sys{} keeps the optimizer outside the trusted base and checks each rewrite, so the optimizer evolves
without re-verification. Translation validation validates each output against its
input~\cite{pnueli1998tv,necula2000tv,zuck2003voc,barrett2005tvoc}, and Alive2 frames the relation as
one-directional refinement for LLVM~\cite{lopes2021alive2}. A validator can itself be proved correct
once and then trusted on every run~\cite{tristan2008validators,tristan2009lcm}, which is the stance
\sys{} takes for its checker. \sys{} formalizes the optimizer-to-checker channel as a certificate
whose claims the checker recomputes, and its refinement is observational under a lens over the
schedules of a cooperative runtime, where the schedule envelope is part of the trusted semantics.

The closest technical neighbors prove local conditions that imply global correctness. The Peek
framework discharges global semantic preservation by checking local properties per assembly
peephole~\cite{mullen2016peek}, and Alive proves rewrites correct at authoring time over sequential
IR~\cite{lopes2015alive}. \sys{} checks each applied rewrite at acceptance time over a concurrent
observation semantics, and its locality is a frame condition over a cooperative scheduler rather than
a liveness fact over a basic block. Credible compilation and witness-passing validators attach
checkable evidence to a transformation~\cite{lerner2003cobalt,rinard1999credible,namjoshi2013witnessing,kang2018crellvm,stepp2011egraphtv};
the certificate of \sys{} plays this role at the source level and is recomputed rather than replayed.
CompCertTSO verifies compilation under relaxed-memory concurrency~\cite{sevcik2013compcerttso}; its
concurrency is preemptive shared memory, while the cooperative scheduling of Scratch is what makes the
cooperative-frame theorem sound and far less conservative than a preemptive analysis allows. The
certificate-carrying stance descends from proof-carrying code~\cite{necula1997pcc}, and like a small
verified kernel~\cite{klein2009sel4} its guarantee rests on a trusted base a reviewer can audit.

Event-driven execution relates Scratch to synchronous reactive languages, which have verified
compilation to a deterministic core~\cite{berry1992esterel,halbwachs1991lustre,bourke2017velus}. Software-engineering work on Scratch~\cite{maloney2010scratch,resnick2009scratch}
studies, analyzes, and transforms its programs: datasets and quality
studies~\cite{aivaloglou2016dataset,aivaloglou2016howkids,robles2017quality}, static analyzers and bug
catalogs~\cite{boe2013hairball,fraedrich2020litterbox,techapalokul2017quality}, and automated test
generation~\cite{stahlbauer2019whisker}. Lens-parametric behavioral equivalence for Scratch
was proposed in recent work~\cite{scratchlens}, on which \sys{} builds certificate-carrying
rewriting and the cooperative-frame theorem.
Recent Scratch-specific work has broadened this stack from static analysis and testing to multimodal feedback, tutoring, benchmarking, repair, assessment, and equivalence: ViScratch uses code and gameplay video for feedback~\cite{si2025viscratch}, Stitch studies stepwise tutoring~\cite{si2025stitch}, ScratchEval packages executable repair tasks and metrics~\cite{si2026scratcheval}, EcoScratch studies cost-aware multimodal repair~\cite{si2026ecoscratch}, Raven uses video-grounded evaluation for assessment~\cite{li2026raven}, ScratchWorld evaluates executable consequences in Scratch worlds~\cite{lin2026scratchworld}, and ScratchLens makes behavioral equivalence lens-parametric~\cite{scratchlens}. Broader programming-education and software-analysis work by Zhang and collaborators studies feedback for competition-level code, LLM-based Python repair, time-limit-exceeded errors, merge-conflict resolution, CI-configuration correctness, and silent misconfiguration detection~\cite{zhang2022clef,zhang2024pydex,zhang2025tle,zhang2022merge,santolucito2022ci,zhang2021configx}. These systems motivate reliable feedback, repair, and analysis pipelines; \sys{} addresses the complementary trust question of accepting source-to-source Scratch transformations only when a small checker can recompute certificate claims.

Closest are two Scratch refactoring systems: quality-driven
refactorings behind editor preconditions~\cite{techapalokul2019refactoring} and search-based readability
transformations validated by generated tests~\cite{adler2021readability}. Both establish behavior
preservation through analysis and testing; Table~\ref{tab:related} places \sys{} against them as the
only one to make each accepted rewrite certificate-carrying, with a recomputing checker and a mechanized
theorem, for this end-user setting~\cite{ko2011eus}. The optimizer follows the
search-based and equality-saturation
tradition~\cite{joshi2002denali,bansal2006superopt,schkufza2013stoke,tate2009egraph,willsey2021egg,nandi2021ruler},
the edits are graph rewrites~\cite{ehrig2006dpo}, and the frame condition adapts separation
reasoning~\cite{reynolds2002separation,ohearn2007resources,abadi1991reasoning} to a cooperative scheduler.

\begin{table}[t]
\centering
\caption{Trust model of Scratch program transformation. Only \sys{} makes each acceptance
certificate-carrying with mechanized soundness.}
\label{tab:related}
\begin{tabular}{@{}lcccc@{}}
\toprule
System & Analysis & Tests & Cert.\ checker & Mech.\ thm. \\
\midrule
QIS~\cite{techapalokul2019refactoring} & \checkmark & & & \\
Search-based~\cite{adler2021readability} & \checkmark & \checkmark & & \\
\sys{} & \checkmark & \checkmark & \checkmark & \checkmark \\
\bottomrule
\end{tabular}
\end{table}

\section{Conclusion}
\label{sec:conclusion}

\sys{} shows that optimization of a concurrent, event-driven, end-user language can carry a
behavior-preservation guarantee: an untrusted optimizer proposes, an auditable checker disposes under
an explicit lens, and the dead-store core is backed by a mechanized cooperative-frame theorem with a
verified-executable Lean realization. On 300 real projects it accepts a rewrite on 283, with no false
accept, and rejects every behavior-changing rewrite a model-free remover would ship.

\balance
\bibliographystyle{IEEEtran}
\bibliography{references}

\end{document}